# SCALING AN ISO COMPLIANCE PRACTICE: STRATEGIC INSIGHTS FROM BUILDING A $1M+ CYBERSECURITY CERTIFICATION LINE

**Nishant Sonkar**
Cisco, Global Cloud Compliance Lead
University of Cincinnati, Ohio
San jose
Nishant19sonkar@gmail.com

**ABSTRACT**

The rapid exponential growth in cloud-first business models and tightened global data protection regulations have led to the exponential increase in the level of importance of ISO certifications, especially ISO/IEC 27001, 27017, and 27018, as strategic imperative propositions for organizations wanting to build trust, ensure compliance, and achieve a competitive advantage. This article describes a case study of a successful design, implementation, and scaling of a cybersecurity certification practice in Armanino LLP, a pioneering US accounting and consulting firm. In reaction to increasing client desires for formalized information security frameworks, I founded an industry practice from conception through implementation to aid mid-market and high-growth technology firms. During one year, the initiative brought in over $ 1 million in new service revenue, expanded our portfolio of cybersecurity clients by 150%, and produced more than 20 successful ISO certifications on various verticals such as SaaS, healthcare, and fintech. Based on the strategic wisdom and operational strategy, this paper outlines the technical architecture of the ISO service line from modular audit templates to certification readiness kits, from stakeholder enablement to integration with SOC 2 and CIS controls. The approach gave value to repeatability, speed, and assurance, thus making Armanino a reputable certification body. The lessons drawn out provide us with a flexible template that can be utilized by firms wishing to build strong compliance programs that can be tailored to address changing digital risk terrains. This work adds to the increasing knowledge about audit scalability, cybersecurity compliance, and ISO standardisation.

1. Introduction

In this day and age of cloud computing and remote working with high investment in data protection regulations, information security frameworks have never been more significant. Among these, ISO/ IEC 27001 and its extensions ISO/ IEC 27017 and ISO/ IEC 27018 have risen to become the fundamental requirement for organizations to protect their asset of information and to comply with the standards of the world. Such certifications strengthen the security posture of a given organization while building the clientele's confidence and creating an opportunity for new business avenues (Feddersen Consulting, 2024).

1.1. The Importance of ISO's certification for Modern businesses

Implementing ISO 27001 and other certifications plays a key role within modern enterprises. These standards give a structured method of controlling the confidential company information, maintaining its confidentiality, integrity, and availability. Allying to ISO 27001 will help organizations discover possible threats, deploy relevant controls, and build a culture for constant improvement in information security management. This proactive approach is essential in curtailing data breaches and meeting the regulatory requirements, like the General Data Protection Regulation (GDPR) (Scytale, 2024).

1.2. Challenges Experienced by the Small to Medium-Sized Firms

Despite clear advantages, adopting and scaling ISO certification practices remains constrained by the serious barriers experienced by the widespread small and medium enterprises (SMEs). The most common challenges are a lack of financial resources, in-house expertise, and the complexity of the certification process. Such limitations can cause problems with adhering to compliance and deriving its maximum value (CCS Risk, 2024). In addition, SMEs may find it challenging to integrate the ISO standards into their current processes without interference in daily operations, thereby complicating the process of certification (DataGuard, 2023).

1.3. **Armanino Strategic Differentiation in a Saturated Market**.

Differentiation becomes very important in a market of firms offering similar services. This need was felt by Armanino LLP, which built its position by establishing a scalable and efficient practice for ISO certification. Although ANAB has enrolled only 29 auditors at this point, Armanino obtained ANAB accreditation for providing ISO/IEC 27001:2022 and ISO/IEC 27701:2019 certification services, which is evidence of Armanino's commitment to maintaining the highest standards for information security and data privacy (Business Wire, 2024). This accreditation affirmed Armanino's abilities and engendered confidence amongst clients seeking credible certification partners.

1.4. **Objective: The construction of a Scalable ISO Practice from the ground up**.

The primary purpose of this article is to explain the strategic wisdom and approaches to realize a scalable ISO compliance practice at Armanino. The article aims to develop a playbook that can be replicated for other organizations by delineating the processes involved, from onboarding clients to creating modular audit templates and interfacing with subsidiary frameworks (such as SOC 2 and CIS). The emphasis here is to form an efficient practice, not only in the sense of being scalable but also in being able to adapt to the changing landscape for practices for clients from different industry sectors.

2. Market Opportunity And Context

The interest in IT Security i) from ISO 27001, ii) ISO 27017, and iii) ISO 27018 certifications has exploded, especially in the domains of technology, fintech, and Software as a Service (SaaS) industries. Demand growth is fueled by the fact that people now require strong information security management systems (ISMS) that will help them protect sensitive data and adhere to global regulations (Junaid, 2023). ISO 27001 has become a major differentiator in the SaaS space. Since the SaaS market is now worth over $195 billion by 2023, companies look forward to presenting a better picture of ensuring data security to show a competitive advantage. ISO 27001 ascertains a comprehensive framework beyond basic security measures that allows SaaS providers to build a strong security posture supporting customer confidence(Kitsios, Chatzidimitriou, & Kamariotou, 2023).

2.1. Limitations of existing solutions (outsourcing, manual auditing)

The fintech companies are increasingly embracing ISO 27001 for data security, with the challenge of stringent regulations. The focus on risk management and continuous improvement of the standard suits the fast-paced nature of the fintech industry. With the incorporation of ISO 27001 with other compliance standards (GDPR and PCI DSS), fintech firms could be able to harmonize their compliance activities and establish trust with stakeholders (Podrecca, Culot, Nassimbeni, & Sartor, 2022). However, it is a pinch point for small to mid-sized firms to scale ISO service lines. Traditional methods, such as outsourcing and manual auditing, can be incredibly resource-intensive and may not give the desired degree of control and customisation. These frailty points highlight the necessity for scalable, efficient, and integrated solutions that can vary depending on the peculiarities of the growing entities (Culot et al., 2021).

**Table 1:** Key compliance drivers further accentuate the importance of ISO certifications.

| Compliance Driver | Description |
| --- | --- |
| Vendor Requirements | Increasingly, clients require their vendors to have ISO certifications to ensure data security. |
| SOC 2 Overlap | ISO 27001 shares standard controls with SOC 2, allowing for streamlined compliance efforts. |
| Cloud Regulation | Regulations like GDPR and HIPAA necessitate robust data protection measures in cloud environments. |

In summary, the escalating demand for ISO certifications in tech, fintech, and SaaS sectors presents a significant market opportunity. Organizations that can develop scalable and efficient ISO compliance practices are well-positioned to meet this demand and differentiate themselves in a competitive landscape.

3. Practice Architecture And Design

3.1. **The design of the E2E ISO Service Delivery Model.**

A strong ISO service delivery model requires a structured approach that includes all dimensions of the certification process. This model should adapt easily to organizational contexts and comply with ISO standards. The design has clearly outlined phases: initial assessment, gap analysis, plans for implementation, control implementation, internal audit, and final certification. All phases should have standardized procedures and documentation to ensure uniformity in operations measured across varied client engagements. Duzenci, Kitapci, and Gok (2023) indicated that organizations' structured, phased approach to compliance is likely to have a high audit success rate and enhanced stakeholder participation. In addition, matching the delivery model with digital transformation strategies like automated controls and controls running on a cloud can also increase scalability and minimize human error in assessments (Donalds & Osei-Bryson, 2020). This alignment is critical in dynamic threat environments (where agility and speed matter).

### 3.2. Audit Preparation Templates

Audit preparation is an integral part of the process for attaining ISO certification. This phase could be streamlined using standardized templates as a standardized procedure and documentation. Free ISO 27001 checklists and templates that can be customized to match specific organizational needs are available from resources such as Smartsheet. These templates usually include a document for the scope of the Information Security Management System (ISMS), risk assessments, control objectives, and evidence for implementing the control. Besides the procedural advantages, audit templates add value in institutional learning knowledge retention, making the process better equipped to withstand staff change and external shocks (Ameen et al., 2021). Templates also guarantee quality and coverage irrespective of industry vertical or client size.

### 3.3. Project Plans and Timelines

Success in the implementation of ISO standards requires effective project administration. In-depth project plan with periods to track the functioning and ensure all tasks are completed on time. For example, the ISO 27001 Implementation Checklist from UpGuard can guide the construction of a complete project plan. These checklists describe the required measures from the first planning stage to final certification and can be modified to suit the varied needs of various companies. Mangundu and Mayayise (2023) advocate for close association of ISO project success with timeline discipline, resource allocation, and clarity of role expectations. Their results also support the process of time-boxed milestone-driven planning in compliance programs.

### 3.4. Certification Checklists

Certification checklists act as guidance for organizations during the ISO certification process. Secureframe offers fully detailed ISO 29001 checklists, including documentation, internal audit, and evidence collection processes during certification. These checklists ensure that no fundamental component is overlooked and that organizations are well prepared for the certification audit. It is demonstrated that the structured use of a checklist significantly reduces the effect on audit preparation time and accuracy, especially when integrated with wider governance, risk, and compliance (GRC) platforms (Mei et al, 2022). Through institutionalizing these tools, firms can guarantee continuity and conformity to fast-changing best practices in cybersecurity compliance.

### 3.5. Use of ISO 27001/27017/27018/27002 Mappings

The combination of several ISO standards necessitates a clear picture of how their controls match and do not match. The relationships of ISO 27001, 27017, 27018, and 27701 controls are mapped in mapping documents such as those provided by NQA. These mappings enable the creation of a unified ISMS with multiple perspectives of information security, cloud security, and privacy. By grasping the overlaps and differences of these standards, organizations can implement controls better and avoid redundant controls.

### 3.6. Creation of Reusable Tools

Creating reusable tools is very important for the consistency and efficiency in the ISO certification process. UnderDefense provides free ISO 27001 policy templates that organizations can follow to adjust to their needs. These templates include a variety of policies such as information security, access control, and incident management. Moreover, standardisation of the training modules and risk assessment templates enables all personnel to be properly prepared and ensures that the risk management processes are uniformly applied throughout the organisation. Such standardization, as Mangundu and Mayayise (2023) describe, should enhance audit outcomes' quality and reproducibility, especially in resource-limited situations. In addition, Mei et al. (2022) highlight that reusable, modular compliance resources alleviate the onboarding load of clients and hasten the time to certification by simplifying the documentation and training workflows.

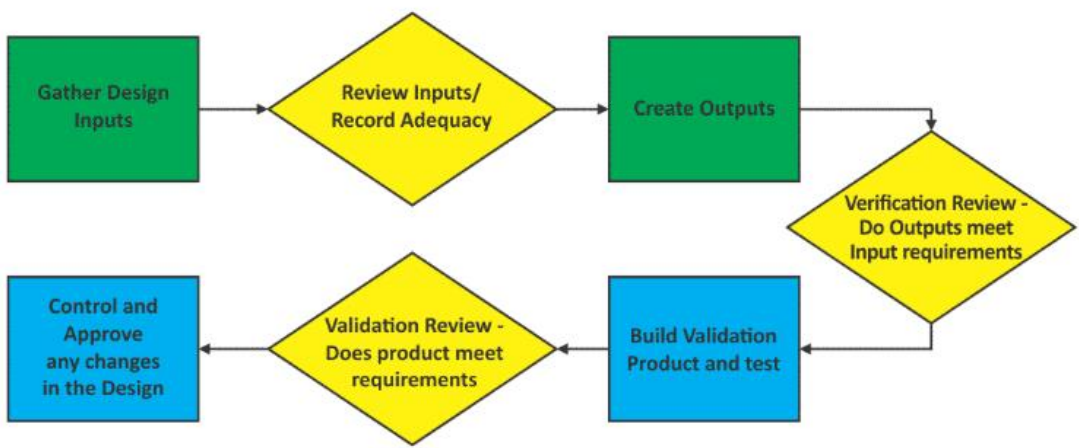

**Figure 1:** ISO Service Delivery Workflow Diagram

This diagram shows the end-to-end process of ISO certification implementation at Armanino, including initial client onboarding, gap assessment, implementation support, internal audits, coordination with the certification bodies, and post-certification continuous improvement loops. The workflow emphasizes modular tool integration and stakeholder involvement, and aligns them with the compliance framework.

## 4. Execution Strategy

### 4.1. First Client Onboarding and Delivery

A strategic attitude towards client onboarding characterized the inception of Armanino's ISO compliance practice; clarity, efficiency, and compliance with ISO 27001 standards are key. The process started with a detailed analysis of the client's existing Information Security Management System (ISMS) and the areas that need improvement and gaps identified. This initial assessment was critical in customizing the certification process to those specific needs of the client, as well as establishing a solid base for compliance. This approach follows pre-established best practices where early-stage diagnostics are necessary to reduce risk and certify readiness (Jakimoski, 2016). Later, a detailed project plan was created to cover scope, aims, schedules, and responsibilities. This plan functioned as a roadmap that outlined the progress of the certification process to the client and Armanino's team. The method also ensured continuous conformity to the risk-based ISO/IEC 27001 framework, ensuring elimination of potential nonconformities and delays (Mei et al., 2022). Regular meetups and progress reviews were introduced to track momentum and the early settling of emerging challenges. Open communication channels and checkpoints facilitated the client's engagement and responsibility.

### 4.2. Playbook for Performing ISO Audits

Armanino's audit methodology was articulated in a structured playbook to standardize the audit approach and deliver client consistency. The playbook outlined every audit step, from planning to reporting, and included best practices and compliance requirements. Incorporating this repeatable framework provided scalability while maintaining the quality and rigor needed to address the influx of clients (Zhao, 2024). The auditor's work had clear phases: scoping the developed system, readiness review, verifying the necessary documentation, control validation, and final certificate decision. Armanino capitalized on data-driven tools and templates at every stage to help drive up auditor productivity and minimize human error. This audit standardization reflects the research results that indicate that structured compliance workflows increase the efficiency and reliability of information assurance activities (Song, Shi, Fischer, & Shankar, 2012). Terralog leads by example when it codifies the ISO audit journey into a playbook and incorporates feedback loops, thus optimizing the delivery model and preparing the ground for sustained service refinement. The key plays for the playbook were:

#### 4.2.1. Audit Planning
The audit scope, objectives, and criteria; selection of competent auditors.

#### 4.2.2. Document Review
Reviewing the client's ISMS documentation to confirm that this matches the ISO 27001 requirement.

#### 4.2.3. Fieldwork
Perusing services through conducting interviews, tests, and observation to evaluate the implementation as well as the effectiveness of controls.

#### 4.2.4. Reporting
Recording findings, including non-conformities and improvement opportunities, and making recommendations. This systematic approach helps conduct thorough objective audits, contributing to the certification process's credibility and reliability.

### 4.3. Stakeholder Training and Handholding in Certification

The importance of personnel as a crucial part of gaining and sustaining ISO 27001 certification has led Armanino to develop extensive training programs for stakeholders across all levels. These programs were intended to increase awareness, build competence levels, and create a culture of information security in client organizations. Training initiatives encompassed:

#### 4.3.1. Awareness Sessions
Training the employees about the importance of information security and their roles in protecting data.

#### 4.3.2. Role-Specific Training
Training on a need-to-know basis for persons with defined roles in the ISMS, such as risk assessment or incident response.

#### 4.3.3. Executive Briefings
The top management should be involved in gaining commitment and support for the ISMS.

#### 4.3.4. These efforts were targeted at ISO 27001 Clause 7.2:
An organization shall ensure that personnel are competent concerning the knowledge, skills, and experience required to perform their function.

**4.4. Collaboration with Certification Bodies (Armanino as CB)**

In the role of a Certification Body (CB), Armanino took on a dual role in the certification process, guiding the clients through the preparation and performing the certification audits. This dual ability required rigorous adherence to impartiality and/or non-objectivity to ensure the certification process's integrity. Contact with clients included communication of audit procedures, criteria, and expectations. Armanino made sure that clients were well informed about the certification process, and also the two-stage audit process:

**4.4.1 Stage 1 Audit**
The client's ISMS document will be inspected before the full audit to determine whether it is ready.

**4.4.2. Stage 2 Audit**
Comprehensive review of the implementation and effectiveness of the ISMS, including on-site review.

This structuring method promoted a transparent and time-saving certification process, strengthening Armanino's position as a trusted CB.

**Table 2:** ISO 27001 Audit Process Overview

| Audit Stage | Description | Objectives |
| --- | --- | --- |
| Stage 1 | Documentation Review | Assess readiness and identify areas for improvement |
| Stage 2 | Implementation Assessment | Evaluate the effectiveness of ISMS and compliance with ISO 27001 |
| Surveillance Audits | Periodic reviews post-certification | Ensure ongoing compliance and continual improvement |
| Recertification Audit | Comprehensive review every three years | Confirm sustained conformity and effectiveness of ISMS |

By employing such an approach to planning and execution, the key point of the implementation will be focused on a structured, collaborative, and transparent approach to TTC's certification with ISO 27001. After incorporating detailed planning, full training, and strict auditing practices, Armanino achieved an effective, scalable, and credible ISO-compliant practice.

**4.5. Measurable Outcomes and Industry Influence**
The tactical development and deployment of Armanino's ISO-certified practice had considerable tangible results, highlighting the firm's role in regulated industries. This section outlines the quantitative and qualitative outcomes secured in the first year since the practice began.

**Table 3:** Key performance indicators of the ISO certification practice of Armanino

| Metric | Outcome |
| --- | --- |
| Revenue Generated | Over $1 million |
| Certified Clients | 20+ organizations |
| Client Base Growth | 150% increase |

Such metrics show the practice's rapid expansion and the market's tendency to seek heavy cybersecurity compliance solutions. The lucrative revenue generation indicates the practice's financial viability and the client's value of getting certified ISO.

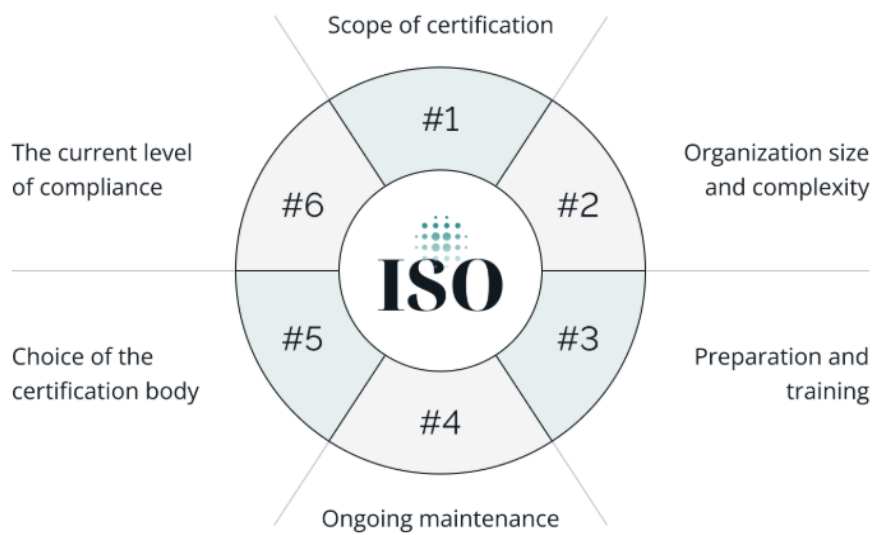

**Figure 2:** Growth indicators highlight the scalability and financial viability of Armanino's ISO certification practice, reflecting strong market demand for cybersecurity compliance solutions.

**4.5.1. Industry Influence**
The ISO certification services of Armanino have permeated many regulated industries deeply:

**4.5.2. SaaS**
The volume of users' data that SaaS companies process. These companies have used Armanino's certifications to assure the customers of their data security, leading them to have a competitive advantage.

**4.5.3. Retail**
ISO certifications have aided retailers in obtaining customer payment information, meeting data protection rules, and minimizing data breaches.

**4.5.4. Healthcare**
Healthcare organizations handling sensitive patient information have adopted certifications to comply with health data regulations and protect patient privacy.

The firm's ability to support these heterogeneous industries reflects the versatility and applicability of the ISO certification services.

**4.6. Transformation into a Trusted Certification Body**
Armanino's commitment to quality in cybersecurity compliance was finalized through accreditation status of the ANSI National Accreditation Board (ANAB) for issuing the ISO/IEC 27001:2022 and ISO/IEC 27701:2019 certifications. This certification means the firm follows international standards and can offer credible and trustworthy certification services. As a press release indicates, this accomplishment sets Armanino as a leading option for organizations needing to demonstrate the highest levels of information security and data privacy.

**4.7. Audit Evidence from Audit Records and Client Success**
Internal audit records and client testimonials of Armanino further attest to its successful practice for ISO certification. Clients have testified to improved operational efficiency and better risk management after accreditation. For example, Stephen Powell noted that lessons learned from Armanino's ISO 27001 Internal Audit and Privacy advisory services helped them mature their ISO Control Framework and develop a Privacy Control Framework fit for several global privacy regulations.

**DIFFERENTIATORS AND INNOVATIONS**

**5.1. Scalable Practice Architecture: Pronto Modular Templates and Audit Readiness Kits**
Development of modular templates and audit readiness kits is one of the cornerstones of Armanino's scalable ISO certification practice. Such resources make the certification process standardized, as it can be applied to different clients and industries. The templates include policy documentation, risk analysis mechanisms, and control matrices specially formulated to match the ISO 27001, 27017, and 27018 standards. According to Song, Shi, Fischer, and Shankar (2012), standardized frameworks increase process efficiency, and service providers can expand their scale without quality loss. Through these pre-structured materials to clients, Armanino saves the time and resources needed for certification, increasing efficiency and scalability.

**5.2. ISO Mapping Accelerators**
Armanino utilizes ISO mapping accelerators to simplify the certification's performance further. These tools align the ISO standards with other compliance systems, such as SOC 2 and CIS Controls. Jakimoski (2016) observes that such integrated tools enhance the speed and accuracy of risk identification, but at the same time, reduce overlaps in frameworks. By recognizing shared requirements and controls, the accelerators permit a synergistic approach to compliance that eliminates redundancy and allows for a complete corporate service. This integrated approach eases the certification process and improves the organization's security posture.

**5.3. Strategic Integration of Compliance Frameworks**
Strategic integration of various compliance frameworks is part of Armanino's innovative approach to enable continuous improvement. By integrating ISO standards with SOC 2 and CIS Controls, the firm develops an effective compliance ecosystem that considers different aspects of information protection and data privacy. Zhao (2024) points out that the convergence of standards and automated control frameworks increases the degree of audit maturity and adaptability, particularly in the dynamic regulatory environment. This comprehensive approach makes clients sufficiently capable of complying with various regulatory requirements and being flexible concerning the changing security challenges.

**Table 4:** Strategic Integration of Compliance Frameworks

| Framework | Focus Area | Integration Benefit |
|---|---|---|
| ISO 27001 | Information Security Management | Establishes a systematic approach to managing sensitive information |
| SOC 2 | Service Organization Controls | Assures the controls relevant to security, availability, processing integrity, confidentiality, and privacy |
| CIS Controls | Cybersecurity Best Practices | Offers a set of prioritized actions to protect organizations from cyber threats |

Lessons learnt from the establishment of Armanino as an ISO certification body. Such a transformation of Armanino to be an accredited ISO certification body provides a valuable learning experience in developing an organization and service differentiation. Key takeaways include:

**5.3.1. Commitment to Impartiality**
Being objective when making certification decisions is important. Armanino set up an impartiality committee to monitor assessment activities and provide impartial results.

**5.3.2. Robust Governance Structures**
Establishing clear governance policies/procedures helps ensure a disciplined and transparent certification process.

**5.3.3. Continuous Improvement**
Continuous updating of certification methodologies and tools regarding evolving threats and regulatory rules guarantees the relevance and impact of the certification services.

**5.3.4. Client Education and Support**
Offering clients access to complete advice and resources facilitates a collaborative orientation toward certification that increases client satisfaction and success rate.

By concentrating on these areas, Armanino not only gained accreditation but also enjoyed a reputation for excellence and reliability in the field of ISO certification.

**Conclusion**

The successful rollout of an arbitrary, high-impact ISO-compliant practice at Armanino is a textbook example of strategic execution, operational discipline, and market responsiveness. Designed around modular templates, built with ISO-SOC-CIS mapping accelerators that plug into the continuous improvement mechanisms proactively at each stage of the certification lifecycle, the practice not only added operational scale, but your client impact was profound. The outcome (over $1 million annual revenue, 150% growth in client base, and 20 + ISO certifications) confirms the power of integration of technical standardization and service innovation. Armanino's transition to becoming an ISO certification body certified was no less important; it was an important step that added to the credibility, maintained impartiality, and provided a new revenue stream based on trust and technical competence. The ISO compliance landscape is moving toward the future because of the introduction of technology and increased regulatory complexities. Artificial intelligence (AI) and machine learning will be increasingly incorporated into audit and certification practices and are capable of real-time risk evaluation analysis, automated control testing, and predictive compliance modelling. Not only will this speed up audit timelines, but it will also minimize manual error and enhance the objectivity of all findings. Also, GRC platforms merging with the ISO framework will enable the centralized management of compliance, continuous monitoring, and reporting for multiple regulatory regimes.

While opportunities abound for mid-market firms that want to scale their ISO offerings, it can be understood that various strategic imperatives are present. First, this should involve investment in standardized and reusable toolkits that eliminate onboarding friction and introduce repeatable delivery models. Second, embrace an integrated mindset of compliance – put frameworks that will support ISO alongside SOC 2, HIPAA, and GDPR, among others. Third, commit to educating stationery in your internal teams and your clients, ensuring compliance is approached not as a routine exercise taught as a checkbox but one shared where the value is added. Finally, pursue accreditation where appropriate; being a certification body can create a meaningful difference for a firm in a crowded market. Adopting these principles will allow companies to scale safely, deliver tangible value, and set themselves at the forefront of a dynamic compliance ecosystem where credibility, automation, and strategic alignment determine long-term success.

# REFERENCES

[1] Feddersen Consulting. (2024). Why ISO 27001 Certification is Critical for Modern Businesses. Retrieved from https://www.feddersenconsulting.com/iso-27001-certification-importance/

[2] Scytale. (2024). Who Needs ISO 27001 Certification? Retrieved from https://scytale.ai/resources/who-needs-iso-27001-certification/

[3] CCS Risk. (2024). What Hinders Businesses from Gaining Value from ISO Standards? Retrieved from https://www.ccsrisk.com/stopping-iso-value

[4] DataGuard. (2023). ISO 27001 for Small Businesses - A Detailed Guide. Retrieved from https://www.dataguard.com/blog/iso-27001-for-small-businesses-a-detailed-guide

[5] Business Wire. (2024). Armanino Certified LLC Achieves ANAB Accreditation for ISO/IEC 27001:2022 and ISO/IEC 27701:2019 Certification Services. Retrieved from https://www.businesswire.com/news/home/20240730963104/en/Armanino-Certified-LLC-Achieves-ANAB-Accreditation-for-ISOIEC-270012022-and-ISOIEC-277012019-Certification-Services

[6] Kitsios, F., Chatzidimitriou, E., & Kamariotou, M. (2023). The ISO/IEC 27001 Information Security Management Standard: How to Extract Value from Data in the IT Sector. Sustainability (Switzerland), 15(7). https://doi.org/10.3390/su15075828

[7] Culot, G., Nassimbeni, G., Podrecca, M., & Sartor, M. (2021, March 16). The ISO/IEC 27001 information security management standard: literature review and theory-based research agenda. TQM Journal. Emerald Group Holdings Ltd. https://doi.org/10.1108/TQM-09-2020-0202

[8] Junaid, T.-S. (2023). ISO 27001: Information Security Management Systems. ResearchGate, 4. Retrieved from https://www.researchgate.net/publication/367166657

[9] Podrecca, M., Culot, G., Nassimbeni, G., & Sartor, M. (2022). Information security and value creation: The performance implications of ISO/IEC 27001. Computers in Industry, 142. https://doi.org/10.1016/j.compind.2022.103744

[10] Duzenci, A., Kitapci, H., & Gok, M. S. (2023). The Role of Decision-Making Styles in Shaping Cybersecurity Compliance Behavior. Applied Sciences (Switzerland), 13(15). https://doi.org/10.3390/app13158731

[11] Donalds, C., & Osei-Bryson, K. M. (2020). Cybersecurity compliance behavior: Exploring the influences of individual decision style and other antecedents. International Journal of Information Management, 51. https://doi.org/10.1016/j.ijinfomgt.2019.102056

[12] Ameen, N., Tarhini, A., Shah, M. H., Madichie, N., Paul, J., & Choudrie, J. (2021). Keeping customers' data secure: A cross-cultural study of cybersecurity compliance among the Gen-Mobile workforce. Computers in Human Behavior, 114. https://doi.org/10.1016/j.chb.2020.106531

[13] Mangundu, J., & Mayayise, T. (2023). The impact of technostress creators on academics' cybersecurity fatigue in South Africa. Issues in Information Systems, 24(4), 294–310. https://doi.org/10.48009/4_iis_2023_123

[14] Ma, Y., Liu, Y., Appolloni, A., & Liu, J. (2021). Does green public procurement encourage firms' environmental certification practices? The mediation role of top management support. Corporate Social Responsibility and Environmental Management, 28(3), 1002–1017. https://doi.org/10.1002/csr.2101

[15] Söderman, M., Wennman-Larsen, A., Hoving, J. L., Alexanderson, K., & Friberg, E. (2022). Interventions regarding physicians' sickness certification practice–a systematic literature review with meta-analyses. Scandinavian Journal of Primary Health Care, 40(1), 104–114. https://doi.org/10.1080/02813432.2022.2036420

[16] Chokhani, S., Ford, W., Sabett, R., Merrill, C., & Wu, S. (2003). Internet X.509 Public Key Infrastructure Certificate Policy and Certification Practices Framework. Network Working Group, IETF, (3647), 94. Retrieved from https://datatracker.ietf.org/doc/rfc3647/

[17] Othman, B., Shaarani, S. M., & Bahron, A. (2016, March 1). The potential of ASEAN in halal certification implementation: A review. Pertanika Journal of Social Sciences and Humanities. Universiti Putra Malaysia.

[18] Boiral, O., & Gendron, Y. (2011). Sustainable Development and Certification Practices: Lessons Learned and Prospects. Business Strategy and the Environment, 20(5), 331–347. https://doi.org/10.1002/bse.701

[19] Mei, R., Yan, H. B., He, Y., Wang, Q., Zhu, S., & Wen, W. (2022). Considerations on Evaluation of Practical Cloud Data Protection. In Communications in Computer and Information Science (Vol. 1699 CCIS, pp. 51–69). Springer Science and Business Media Deutschland GmbH. https://doi.org/10.1007/978-981-19-8285-9_4

[20] Song, D., Shi, E., Fischer, I., & Shankar, U. (2012). Cloud data protection for the masses. Computer, 45(1), 39–45. https://doi.org/10.1109/MC.2012.1

[21] Zhao, N. (2024). Improvement of Cloud Computing Medical Data Protection Technology Based on Symmetric Encryption Algorithm. Journal of Testing and Evaluation, 52(1). https://doi.org/10.1520/JTE2021-0456

[22] Jakimoski, K. (2016). Security techniques for data protection in cloud computing. International Journal of Grid and Distributed Computing, 9(1), 49–56. https://doi.org/10.14257/ijgdc.2016.9.1.05

[23] Ahmad, A., Alabduljabbar, A., Saad, M., Nyang, D., Kim, J., & Mohaisen, D. (2021). Empirically comparing the performance of blockchain's consensus algorithms. IET Blockchain, 1(1), 56–64. https://doi.org/10.1049/blc2.12007

[24] Carden, L., Maldonado, T., Brace, C., & Myers, M. (2020). WITHDRAWN – Administrative Duplicate Publication: Robotics process automation at TECHSERV: An implementation case study. Journal of Information Technology, 35(1), NP1–NP9. https://doi.org/10.1177/0268396219881455

[25] Wang, M., Wu, Q., Qin, B., Wang, Q., Liu, J., & Guan, Z. (2018). Lightweight and Manageable Digital Evidence Preservation System on Bitcoin. Journal of Computer Science and Technology, 33(3), 568–586. https://doi.org/10.1007/s11390-018-1841-4

[26] Deswarte, Y. (2011). Protecting critical infrastructures while preserving each organization's autonomy. In Lecture Notes in Computer Science (including subseries Lecture Notes in Artificial Intelligence and Bioinformatics) (Vol. 6536 LNCS, pp. 15–34). https://doi.org/10.1007/978-3-642-19056-8_2

[27] Ar, I. M., Erol, I., Peker, I., Ozdemir, A. I., Medeni, T. D., & Medeni, I. T. (2020). Evaluating the feasibility of blockchain in logistics operations: A decision framework. Expert Systems with Applications, 158. https://doi.org/10.1016/j.eswa.2020.113543